\documentstyle[12pt]{article}

\def\lappeq{\mathrel{\rlap{\raise.5ex\hbox{$<$}}{\lower.5ex\hbox{$\sim$}}}}

\begin{document}

\begin{flushright}
{ TO APPEAR IN {\it NATURE}}\\

{ (UNDER PRESS EMBARGO UNTIL PUBLICATION)}\\
\end{flushright}

\begin{centering}
\bigskip
{\Large {\bf Gravity-wave interferometers as\\
quantum-gravity detectors} }

(title may be changed in the journal)
\bigskip
\bigskip

{\bf Giovanni AMELINO-CAMELIA}\\
\bigskip
Institut de Physique, Universit\'e de Neuch\^atel,
     CH-2000 Neuch\^atel, Switzerland
\end{centering}
\vspace{1cm}

\baselineskip = 12pt

\noindent

{\bf A prediction of nearly
all approaches to the unification
of Quantum Mechanics and Gravity is that
at very short distances the sharp
classical concept of space-time should give way
to a somewhat ``fuzzy'' (or ``foamy'')
%picture~\cite{wheely,arsarea,emn,hawk}.
picture~[1-4].
The properties of this fuzziness and the length scale
characterizing its onset are potentially
a means for determining which (if any) of the existing Quantum
Gravity models is correct, but
it has been generally believed \cite{nodata}
that the smallness of the
quantum space-time effects would not allow to study them
with presently available technologies. Here I observe that
some proposals for the nature of this space-time fuzziness
would affect the operation
of gravity-wave interferometers
by effectively introducing an additional
source of noise that can be tightly constrained
experimentally.
In particular, I show that noise levels recently achieved
rule out values of the length scale that characterizes
one of the fuzziness proposals down to
the Planck length ($L_p \! \sim \! 10^{-35}m$) and beyond,
while for another proposal Planck-level sensitivity
is within reach of
gravity-wave interferometers
that will start operating in the near future.}

In a fuzzy space-time the operative
definition of a distance $D$
is affected by quantum fluctuations. These fluctuations
are primarily characterized by their overall
magnitude $\sigma_D$ (the root-mean-square deviation of $D$).
The simplest proposals are such that
\begin{equation}
\sigma_D \ge L_{min} \, ,
\label{no1}
\end{equation}
where $L_{min}$ is
a Quantum-Gravity scale expected to be simply
related to (usually identified with) the Planck length.
Relations of the type (\ref{no1}) are motivated by
certain analyses of gedanken experiments (see, {\it e.g.},
Ref.~\cite{garay})
that combine some elements of Quantum Mechanics and
some elements of Gravity.
The quantum space-time fluctuations
responsible for (\ref{no1})
are often visualized
as involving geometry and topology fluctuations~\cite{wheely},
virtual black holes~\cite{hawk},
and other novel phenomena.

Other scenarios for space-time fuzziness
arise when taking into account the quantum
properties of devices, which were ignored
in the original studies~\cite{garay}
that led to the proposal of Eq.~(\ref{no1}).
It is well understood
(see, {\it e.g.}
%Refs.~\cite{bergstac,diosi,ng,dharam94grf,gacmpla,grf98})
Refs.~[7-12])
that the combination of
the gravitational properties
and
the quantum
properties
of devices can have an important
role in the analysis
of the operative definition of gravitational observables.
The implications can be far-reaching; in particular,
in Ref.~\cite{dharam94grf} it was observed that
the masses of the probes used in measurements induce
a change in the space-time metric and this
is associated to the emergence of nonlocality.
The nature of the
gravitationally-induced nonlocality
suggests \cite{dharam94grf} a modification of the
fundamental commutators.

Here I shall be primarily concerned with the role
that the quantum and the gravitational properties
of devices have in the
analysis of the measurability of distances, in the sense
first illustrated in an influential study
by Wigner~\cite{wign}.
Wigner derived a quantum limit
on the measurability of the distance $D$
separating two bodies by analysing
a measurement procedure based on the exchange
of a light signal between the bodies.
Taking into account Heisenberg's position-momentum
uncertainty relations also for the clock used
in the measurement procedure
Wigner obtained a lower bound on the
quantum uncertainty in $D$:
\begin{eqnarray}
\delta D \ge  \sqrt{{\hbar T_{obs} \over 2 M_c }}
\sim  \sqrt{{\hbar D \over c M_c }}
~,
\label{deltawignOLD}
\end{eqnarray}
where $M_c$ is the mass of the
clock, $T_{obs}$ is the time required by the measurement
procedure, and on the right-hand-side I used the
fact that the Wigner measurement
of a distance $D$ requires a time $2 D/c$.

The result (\ref{deltawignOLD}) may at first
appear somewhat puzzling, since
ordinary Quantum Mechanics should not
limit the measurability of any given observable.
[It only limits the combined measurability
of pairs of conjugate observables.]
However, Quantum Mechanics is the theoretical
framework for the description of the outcome of experiments
performed by classical devices.
In the limit in which the
devices ({\it e.g.} Wigner's clock)
behave ``classically'', which in particular requires
the devices to be infinitely massive
(so that $\delta x \, \delta v \sim \hbar/m \sim 0$),
the right-hand side of equation (\ref{deltawignOLD}) tends
to zero.
Therefore, as expected, there is
no limitation on the measurability
of the distance $D$ in the
appropriate infinite-mass ``classical-device limit.''
This line of argument depends crucially on the fact that
ordinary Quantum Mechanics does not involve Gravity.

Quite clearly the classical infinite-mass limit is not consistent
with the nature of measurements involving gravitational effects.
As the devices get more and more massive they increasingly
disturb the gravitational/geometrical observables, and
well before reaching the infinite-mass limit the procedures
for the measurement of gravitational observables cannot
be meaningfully performed~\cite{ng,gacmpla,grf98}.
A well-known example of this problem has been
encountered in attempts (see, {\it e.g.}, Ref.~\cite{bergstac})
to generalize to
the study of the measurability of gravitational fields
the famous Bohr-Rosenfeld analysis~\cite{rose}
of the measurability of the electromagnetic field.
In order to achieve the accuracy allowed by
the formalism of ordinary Quantum Mechanics,
the Bohr-Rosenfeld measurement procedure resorts to
ideal test particles of infinite mass, which would of course
not be admissable probes in a gravitational context.
Similarly, in Wigner's measurement procedure
the limit $M_c \rightarrow \infty$
is not admissable when gravitational interactions
are taken into account.
At the very least the value of $M_c$ is limited by the
requirement that the clock should not turn into a black hole
(which would not allow the required exchange of signals
between the clock and the other devices).
These observations, which render unavoidable
the $\sqrt{T_{obs}}$-dependence of Eq.~(\ref{deltawignOLD}),
provide motivation for the possibility~\cite{gacmpla,grf98}
that in Quantum Gravity any measurement that monitors a distance $D$
for a time $T_{obs}$ is affected by quantum fluctuations
characterized by
\begin{eqnarray}
\sigma_D  \sim \sqrt{ {L_{QG} \, c \, T_{obs}}}
~,
\label{gacdl}
\end{eqnarray}
where $L_{QG}$ is a fundamental length scale which we can expect
to be simply related to the Planck length.
In particular, the Wigner measurement
of a distance $D$, which requires a time $2 D/c$,
would be affected
by fluctuations of magnitude $\sqrt{L_{QG} D}$.
[Interestingly, the study reported in Ref.~\cite{diosi}
analyzed the interplay of Quantum Mechanics and Gravity
in defining a net of time-like geodesics
and suggested that
the maximum ``tightness'' achievable
for the geodesics net is $\sqrt{L_{QG} D}$.]

A $\sigma_D$ that increases with $T_{obs}$ is not surprising
for space-time fuzziness scenarios; in fact, the same phenomena
that would lead to fuzziness are also expected to
induce ``information loss'' \cite{hawk}
(the information stored in a quantum system
degrades as $T_{obs}$ increases).
The argument based on the Wigner setup provides motivation
to explore the specific form $\sigma_D \! \sim \! \sqrt{T_{obs}}$
of this $T_{obs}$-dependence.

From the type of $T_{obs}$-dependence of Eq.~(\ref{gacdl}),
and the stochastic properties of the
%processes~\cite{wheely,arsarea,emn,hawk}
processes~[1-4]
expected to characterize a fuzzy space-time,
it follows that the quantum fluctuations responsible
for (\ref{gacdl}) should have amplitude spectral density $S(f)$
with the $f^{-1}$ dependence typical of ``random walk noise''~\cite{rwold}:
\begin{eqnarray}
S(f) = f^{-1} \sqrt {L_{QG} \, c} ~.
~
\label{gacspectr}
\end{eqnarray}
If indeed $L_{QG} \sim 10^{-35} m$, from (\ref{gacspectr})
one obtains $S(f) \sim f^{-1} \cdot (5 \cdot 10^{-14} m \sqrt{H\!z})$.
[Of course,
one expects that this formula for the Quantum-Gravity induced $S(f)$
could only apply to frequencies $f$ significantly
smaller than the Planck frequency $c/L_p$
and significantly larger than the inverse of the time scale
over which, even ignoring the gravitational field
generated by the devices, the classical geometry
of the space-time region where
the experiment is performed manifests
significant curvature effects.]

Before commenting on how the proposal (\ref{gacdl})-(\ref{gacspectr})
compares with data from modern gravity-wave interferometers,
let me consider another
space-time fuzziness scenario, which involves
fluctuations of significantly different magnitude.
This alternative scenario~\cite{ng,karo} is
essentially based on the observation that
the uncertainty described by
Wigner's Eq.~(\ref{deltawignOLD}) can be combined with
a classical-Gravity estimate
of the uncertainty in the measurement of the distance $D$
that results from the distorsion of geometry associated
to the gravitational field generated by the clock.
While the uncertainty (\ref{deltawignOLD}) dicreases with $M_c$,
the uncertainty induced by Gravity increases with $M_c$,
and combining the two uncertainties
one finds a minimum total uncertainty
of the type $\delta D  \sim ({\cal L}_{QG}^2 \, c \, T_{obs})^{1/3}$,
where ${\cal L}_{QG}$ is a length scale
analogous to $L_{QG}$.
Just like in the other analyses
of the dependence of the Wigner measurement
on the time of observation,
one is then led to consider a fuzzy space-time
with corresponding $T_{obs}$-dependence:
\begin{eqnarray}
\sigma_D  \sim ({\cal L}_{QG}^2 \, c \, T_{obs})^{1/3}
~.
\label{ngdl}
\end{eqnarray}
The associated amplitude spectral density is
\begin{eqnarray}
{\cal S}(f) = f^{-5/6} ({\cal L}_{QG}^2 \, c)^{1/3}
~,
\label{ngspectr}
\end{eqnarray}
which for ${\cal L}_{QG} \sim 10^{-35} m$
gives ${\cal S}(f) = f^{-5/6} \cdot (3 \cdot 10^{-21} m \, H\!z^{1/3})$.

Each of the proposals
(\ref{no1}),
(\ref{gacdl}), (\ref{ngdl})
was obtained within
a corresponding scheme for the interplay between
Quantum Mechanics and weak-field Gravity.
This is an approach that has already proven successful
in Quantum-Gravity research; in fact,
the phenomenon of gravitationally induced
phases, which was also predicted from analyses
of the interplay between
Quantum Mechanics and weak-field Gravity,
has already been confirmed experimentally~\cite{cow}.
By discovering experimentally
which of the space-time fuzziness
proposals is correct,
we could also obtain additional insight in the
weak-field limit of Quantum Gravity,
and the requirement of
consistency with the correct weak-field limit
can represent a highly non-trivial constraint
for the search of Quantum Gravity.
In particular, the very
popular Quantum-Gravity theories
based on {\it Critical Strings} appear to require
fuzziness of type (\ref{no1}), as seen in
analyses of string collisions at Planckian
energies~\cite{amaciaven}, and a proposal for a quantum-group
structure which might accommodate (\ref{no1}) has been
discussed in Ref.~\cite{kempf}.
The proposal (\ref{gacdl})
has been found to arise within the mathematical framework
of dimensionfully deformed
Poincar\'e symmetries\cite{kpoin,gacxt}, which has been
attracting much interest recently.
Point-particle Quantum-Gravity theories
based on these deformations would
therefore require
space-time fuzziness of type (\ref{gacdl}).
Moreover, Eq.~(\ref{gacdl})
has been shown to hold within Liouville
({\it non-critical}) String Theory~\cite{emn,aemn1},
another approach to Quantum Gravity which is attracting
significant interest.
The search of Quantum-Gravity theories whose weak-field
limit is consistent with (\ref{ngdl}) has not yet been
successful, but there appears to be no in principle
obstruction and therefore one can expect
progress in this direction to be forthcoming.

While conceptually
the proposals (\ref{no1}),
(\ref{gacdl}) and (\ref{ngdl})
represent drastic departures from conventional physics,
phenomenologically they appear to encode only minute effects;
for example, it has been observed that,
assuming $L_{min}$, $L_{QG}$ and ${\cal L}_{QG}$  are not much
larger than the Planck length,
all of these proposals encode submeter fluctuations on
the size of the whole observable universe
(about $10^{10}$ light years).
However, the precision~\cite{saulson}
of modern gravity-wave interferometers is such that
they can provide significant information at least on
the proposals (\ref{gacdl}) and (\ref{ngdl}).
In fact, the operation of gravity-wave interferometers
is based on the detection of
minute changes in the positions of some test masses (relative to
the position of a beam splitter).
If these positions were
affected by quantum fluctuations of the type discussed
above the operation of gravity-wave interferometers would effectively
involve an additional
source of noise due to Quantum-Gravity.
This observation allows to set interesting bounds already
using existing noise-level data obtained at
the {\it Caltech 40-meter interferometer}.
This interferometer has achieved~\cite{ligoprototype}
displacement noise levels with amplitude spectral density
lower than $10^{-18} m/\sqrt{H\!z}$
for frequencies between $200$ and $2000$ $H\!z$
and this, as seen by straightforward comparison
with Eq.~(\ref{gacspectr}), clearly rules out all values of $L_{QG}$
down to the Planck length.
Actually, even values of $L_{QG}$ significantly
lower than the Planck length are inconsistent with the data reported
in Ref.~\cite{ligoprototype}; in particular,
by confronting Eq.~(\ref{gacspectr}) with the observed
noise level of $3 \cdot 10^{-19} m/\sqrt{H\!z}$ near $450$ $H\!z$,
which is the best achieved at the {\it Caltech 40-meter interferometer},
one obtains the bound $L_{QG} \le 10^{-40}m$.
While at present we should allow for some relatively small factor
to intervene in the relation between $L_{QG}$ and $L_p$,
having excluded all values of $L_{QG}$
down to $10^{-40}m$ the status of the proposal (\ref{gacdl})
appears to be at best problematic.
Of course, even more stringent bounds on $L_{QG}$
are within reach of the next LIGO/VIRGO~\cite{ligo,virgo}
generation of gravity-wave interferometers.

The sensitivity achieved at
the {\it Caltech 40-meter interferometer}
also sets a bound on the proposal (\ref{ngdl})-(\ref{ngspectr}).
By observing that Eq.~(\ref{ngspectr}) would imply Quantum-Gravity
noise levels for gravity-wave interferometers of
order ${\cal L}_{QG}^{2/3} \cdot (10 \, m^{1/3}/\sqrt{H\!z})$
at frequencies of a few hundred $H\!z$, one obtains
from the data reported in Ref.~\cite{ligoprototype}
that ${{\cal L}_{QG}} \le 10^{-29} m$.
This bound is remarkably stringent in absolute terms, but is still
well above the range of values of ${{\cal L}_{QG}}$ that is
favoured by the intuition emerging
from the various theoretical approaches to Quantum Gravity.
A more significant bound on ${{\cal L}_{QG}}$
should be obtained by the LIGO/VIRGO generation
of gravity-wave interferometers.
For example, it is plausible~\cite{ligo} that
the ``advanced phase'' of LIGO achieve a displacement noise spectrum
of less than $10^{-20} m/ \sqrt{H\!z}$ near $100$ $H\!z$
and this would probe values of ${{\cal L}_{QG}}$ as small
as $10^{-34} m$.

Looking beyond the LIGO/VIRGO generation
of gravity-wave interferometers, one can envisage still quite
sizeable margins for improvement by optimizing the
performance of the interferometers at low frequencies,
where both (\ref{gacspectr}) and (\ref{ngspectr}) become
more significant. It appears natural to perform such studies
in the quiet environment of space, perhaps through future
refinements of LISA-type setups~\cite{lisa}.

The example of gravity-wave interferometers here emphasized
shows that the smallness of the Planck length
does not preclude the possibility
of direct investigations of space-time fuzziness.
This complements the results of the studies~\cite{ehns,grbgac}
which had shown that indirect evidence of quantum space-time
fluctuations could be obtained by testing the predictions of
theories consistent with a given picture of these fluctuations.
Additional encouragement for the outlook of experimentally-driven
progress in the understanding of the interplay between
Gravity and Quantum Mechanics comes from recent
studies~\cite{dharamwin,atomic}
in the area of gravitationally induced phases,
whose significance has
been emphasized in Refs.~\cite{sci,newsci}.

\baselineskip 12pt plus .5pt minus .5pt

\baselineskip = 12pt

\section*{Acknowledgements}
I owe special thanks to Abhay Ashtekar, since he
suggested to me that gravity-wave interferometers might be
useful for experimental tests of some of the Quantum-Gravity
phenomena that I have been investigating.
Still on the theoretical side I am grateful to several colleagues
who provided encouragement and stimulating feed-back,
particularly D.~Ahluwalia, J.~Ellis, J.~Lukierski,
N.E.~Mavromatos, C.~Rovelli, S.~Sarkar and J.~Stachel.
On the experimental side I would like to thank
F.~Barone,
M.~Coles,
J.~Faist,
R.~Flaminio,
L.~Gammaitoni,
G.~Gonzalez,
T.~Huffman,
L.~Marrucci
and
M.~Punturo
for useful conversations
on various aspects of interferometry.
I acknowledge support by a grant of
the Swiss National Science Foundation.

\end{document}